\documentclass[12pt,letterpaper]{article}
\usepackage[utf8]{inputenc}

\pdfoutput=1
\usepackage{color}
\definecolor{darkblue}{rgb}{0.1,0.1,.7}
\usepackage[colorlinks, linkcolor=darkblue, citecolor=darkblue, urlcolor=darkblue, linktocpage]{hyperref}
\usepackage[]{amsmath}
\usepackage[]{graphicx}
\usepackage[]{latexsym}
\usepackage[utf8]{inputenc}
\usepackage{slashed,graphicx,color,amsmath,amssymb}
\usepackage[mathscr]{eucal}
\usepackage{mathrsfs}
\usepackage[margin=10pt,font=small,labelfont=bf]{caption}
\usepackage{amscd}
\usepackage{bm}
\usepackage{xcolor}
\usepackage{upgreek}
\usepackage[square, comma, sort&compress,numbers]{natbib}
\usepackage[all,cmtip]{xy}
\usepackage[margin=1.7in]{geometry}
\usepackage{cleveref}
\usepackage[color=cyan!30!white,linecolor=red,textsize=footnotesize]{todonotes}
\usepackage{soul}
\numberwithin{equation}{section}
\hypersetup{
	pdfstartview={FitH},    
	pdftitle={Characters of irrelevant deformations},    
	pdfauthor={Datta, Jiang},     
	colorlinks=true,       
	linkcolor=blue,          
	citecolor=blue!59!black! ,        
	filecolor=magenta,      
	urlcolor=blue           
}

\newcommand{\tr}{\mathrm{Tr}\,}

\def\ttb{T{\bar{T}}}
\def\btau{{\bar{\tau}}}

\def\pd{\partial}

\def\bq{{\bar{q}}}

\def\jtbar{J\bar{T}}
\def\bh{{\bar{h}}}

\def\cN{\mathcal{N}}

\def\ie{{\it i.e.~}}

\def\nn{\nonumber}
\def\pd{\partial}

\def\l1{{{1-loop}}}

\def\n1{\Bigg|_{n=1}}

\def\n{{(n)}}
\def\tr{{Tr}}

\def\cN{\mathcal{N}}

\def\tr{\text{Tr}}

\def\bL{\bar{L}}

\def\ein{{(1)}}

\def\bq{\bar{q}}

\def\pd{\partial}

\def\jtbar{J\bar{T}}
\def\bh{{\bar{h}}}

\def\beq{\begin{equation}}
\def\eeq{\end{equation}}

\def\bea{\begin{eqnarray}}
\def\eea{\end{eqnarray}}
\def\nn{\nonumber}
\def\pd{\partial}

\def\l1{{\text{1-loop}}}

\def\n1{\Bigg|_{n=1}}

\def\n{{(n)}}
\def\tr{\text{Tr}}

\def\cN{\mathcal{N}}

\def\cF{\mathcal{F}}
\def\ep{\epsilon}
\def\ein{{(1)}}

\def\tlambda{\widetilde{\lambda}}

\def\tep{\tilde{\epsilon}}

\def\be{\begin{equation}}
\def\ee{\end{equation}}
\def\bal{\begin{array}{l}}
\def\ba#1{\begin{array}{#1}}  
	\def\ea{\end{array}}
\def\bea{\begin{eqnarray}}
\def\eea{\end{eqnarray}}
\def\beas{\begin{eqnarray*}}
	\def\eeas{\end{eqnarray*}}

\def\nn{\\\nonumber}

\def\ttb{{T\bar{T}}}

\def\nn{\nonumber}
\def\bit{\begin{item}}
	\def\eit{\end{item}}
\def\benu{\begin{enumerate}}
	\def\eenu{\end{enumerate}}
\def\tr{{\rm tr}}

\def\tep{\widetilde{\epsilon}}

\def\tilh{\widetilde{h}}

\DeclareFontFamily{U}{wncy}{}
\DeclareFontShape{U}{wncy}{m}{n}{<->wncyr10}{}
\DeclareSymbolFont{mcy}{U}{wncy}{m}{n}
\DeclareMathSymbol{\Sha}{\mathord}{mcy}{"58}

 \makeatletter
 \g@addto@macro\bfseries{\boldmath}
 \makeatother

\makeatletter
\if@todonotes@disabled

\else

\fi
\makeatother

\begin{document}

\definecolor{tinge}{RGB}{255, 244, 195}
\sethlcolor{tinge}
\setstcolor{red}

\vspace*{-.8in} \thispagestyle{empty}
\begin{flushright}
	\texttt{CERN-TH-2021-044}
\end{flushright}
\vspace{.7in} {\Large
\begin{center}
{\LARGE \bf  Characters of irrelevant deformations}
\end{center}}
\vspace{.5in}
\begin{center}
{Shouvik Datta$^1$ \& Yunfeng Jiang$^{1,2}$}
\\
\vspace{.3in}
\small{
  $^1$\textit{Department of Theoretical Physics, CERN,\\
	1 Esplanade des Particules, Geneva 23, CH-1211, Switzerland.}\\
\vspace{.5cm}

$^2$\textit{Shing-Tung Yau Center and School of Physics, Southeast University,\\ Nanjing 210096, China.}\\
\vspace{.5cm}
} \vspace{0cm}
\begingroup\ttfamily\small
\{shouvik.datta,\,yunfeng.jiang\}@cern.ch\par
\endgroup


\end{center}

\vspace{.6in}

\begin{abstract}
\normalsize
We analyse the $T\bar{T}$ deformation of 2d CFTs in a special double-scaling limit, of large central charge and small deformation parameter. In particular, we derive closed formulae for the deformation of the product of left and right moving CFT characters on the torus. It is shown that the $1/c$ contribution takes the same form as that of a CFT, but with rescalings of the modular parameter reflecting a state-dependent change of coordinates. We also extend the analysis for more general deformations that involve $T\bar{T}$, $J\bar{T}$ and $T\bar{J}$ simultaneously. We comment on the implications of our results for holographic proposals of irrelevant deformations.
\end{abstract}

\vskip 1cm \hspace{0.7cm}

\newpage

\setcounter{page}{1}

\noindent\rule{\textwidth}{.1pt}\vspace{-1.2cm}
\begingroup
\hypersetup{linkcolor=black}
\tableofcontents
\endgroup
\noindent\rule{\textwidth}{.2pt}

\section{Introduction}

In 2-dimensional conformal field theories, characters form the building blocks of torus partition functions. For rational CFTs they encode information about null-states, while for irrational models they lead to stringent constraints on the spectrum of primaries once modular invariance is imposed \cite{Cardy:1986ie,Hellerman:2009bu}. In light of recent developments on irrelevant deformations of CFTs, consisting of $\ttb$  and its Lorentz-breaking cousins \cite{Smirnov:2016lqw,Cavaglia:2016oda,Guica:2017lia,Apolo:2018qpq,Chakraborty:2018vja,LeFloch:2019rut,Chakraborty:2019mdf}, it's conceivable that traditional CFT notions of Verma modules and characters generalize in a suitable manner. The goal of this work is to take steps towards realizing this possibility.

The solvable nature of the $\ttb$ and $J\bar T$ deformations hints  that the Hilbert space of the deformed theory has a similar structure to that of the undeformed CFT. It is well-known that in CFTs states organize themselves into irreducible representations of the chiral algebra, with  primaries being the highest weight states and  descendants built from excitations on the primaries. A distinguishing feature of the family of solvable irrelevant deformations is that there is a one-to-one map between the spectra of the deformed and undeformed theories, and the deformed partition functions inherit the modular properties from the undeformed theory. It is therefore reasonable to expect CFT-like organizational principles for the Hilbert space of the deformed theory. Investigating this aspect further has the potential to reveal, a posteriori, how the underlying symmetry algebra (such as, Virasoro) gets deformed.

In this paper we consider $\ttb$ deformations of CFTs at large central charge and small deformation parameter. Specifically, we work in the following double-scaling limit:
\begin{align}
	c\to \infty, \qquad
	\lambda \to 0, \qquad
	c\,\lambda = \text{fixed}~.
\end{align}
This regime is of direct relevance to the holographic proposal of AdS with a radial cutoff \cite{McGough:2016lol,Kraus:2018xrn,Caputa:2020lpa}; the position of the cutoff is related to the $\ttb$ coupling as $\ell_{\rm AdS}^2/r_c^2 = \pi c \lambda /6$.
The following question can then be posed: how does the product of left and right moving CFT characters, $|\chi_h(q)|^2$, get deformed under the $\ttb$ flow? Since the spectrum of the deformed theory is exactly solvable, this question can be readily addressed provided we express the quantity as a sum over Boltzmann factors. The novelty, however, lies in writing product formulae for the characters just like CFT case.

We consider the $\ttb$ deformation of the `free energy', $\log |\chi_h(q)|^2$, in the $1/c$ expansion. In the 3d gravity dual, this expansion corresponds to the loop expansion in Newton's constant, $G_N^{(3)}$. We evaluate the quantity using three methods, which establishes the consistency our results; the methods being : (\emph{i}) using the the explicit form of the deformed spectrum \cite{Smirnov:2016lqw,Cavaglia:2016oda} and performing the $1/c$ expansion; (\emph{ii}) writing the deformed partition function as an integral transformation of the undeformed one \cite{Dubovsky:2017cnj,Dubovsky:2018bmo} and performing a saddle-point analysis in the large $c$ limit; and (\emph{iii}) performing an $1/c$ expansion for the flow equation of the torus partition function \cite{Cardy:2018sdv,Datta:2018thy,Aharony:2018bad}. We find that the deformed character at leading order, $O(c)$, is simply given by the usual solution to the Burger's equation. The $O(c^0)$ contribution strikingly turns out to be of the same form as that of the undeformed case, albeit  rescaling of the imaginary part of the modular parameter ($\tau_2$). The rescaling depends on the $\ttb$ coupling $\lambda$, as well as the conformal dimension and spin the irreducible representation in question. This phenomenon is a consequence of a dynamical, or state-dependent, change of coordinates \cite{Dubovsky:2017cnj,Conti:2018tca,Guica:2019nzm,Caputa:2020lpa}. 

We also generalize our analysis to the $\lambda T\bar{T}+ \epsilon_+ J\bar{T} + \epsilon_- T\bar{J}$ deformation. This joint deformation can be formulated by coupling the undeformed theory to a JT-like topological gravity and a 2d gauge theory \cite{Aguilera-Damia:2019tpe,Chakraborty:2019mdf,Hashimoto:2019wct,Hashimoto:2019hqo,Chakraborty:2020xyz}. The deformed partition function can then be obtained by the action of an integral kernel on the undeformed flavored partition function. Following the analysis for the pure $T\bar{T}$ deformation, we obtain the $O(c)$ and $O(c^0)$ contributions to deformed characters. The structure of result is analogous to the $T\bar{T}$ case, with the one-loop contribution displaying a state-dependent rescaling of  $\tau_2$.

The results for the deformed characters have direct implications for holographic proposals of the irrelevant deformations. Till date, there has been three different holographic proposals for $\ttb$-- cutoff AdS \cite{McGough:2016lol,Kraus:2018xrn,Caputa:2020lpa}, mixed boundary conditions \cite{Guica:2019nzm} and random fluctuating boundaries \cite{Hirano:2020nwq}\footnote{These are proposals for the double-trace version of $\ttb$. See \cite{Giveon:2017myj,Asrat:2017tzd,Chakraborty:2020swe,Chakraborty:2020cgo} for the single-trace version.}. The $1/c$ contribution to the vacuum character corresponds to the 1-loop determinant of the graviton in 3d gravity.
Therefore, the deformed characters
provide sharp predictions for the bulk dual.  We do not undertake this verification as a part of this work and hope to address it in the near future.

This paper is organized as follows. In Section \ref{sec:ttb} we study the deformation of the characters of $\ttb$ deformation using the three approaches outlined above. We then generalize the results to the joint deformations of type $\lambda T\bar{T}+ \epsilon_+ J\bar{T} + \epsilon_- T\bar{J}$ in Section \ref{sec:joint}.  Section \ref{sec:conclusions} contains our conclusions and comments on the implications for holographic constructions.

\textbf{Note added:} While this manuscript was in preparation, reference \cite{Kraus:2021cwf} appeared on the arXiv. Although the focus of this work is different, there is some overlap with Sections 7 and 8 of that paper.


\section{Deformed characters of the $\ttb$ deformation}
\label{sec:ttb}
Let us recall the definition of the $T\bar{T}$ deformation and the deformed spectrum in finite volume. For a QFT described by Lagrangian $\mathcal{L}_0$, the $T\bar{T}$ deformation is defined by \cite{Smirnov:2016lqw,Cavaglia:2016oda}
\begin{align}
	\frac{dS}{d\mu}=\int\det \left[T_{ab}^{(\mu)}(x)\right]\,  d^2x~,
\end{align}
where $\mu$ is the deformation parameter and $S$ is the deformed action. We consider the theory on a cylinder of radius $R$. For a given energy eigenstate $|n\rangle$, we denote the corresponding energy and momentum by $E_n(R)$ and $P_n(R)$ respectively. The deformed energy ${E}_n(R,\mu)$ satisfies the following flow equation
\begin{align}
	\label{eq:flowE}
	\partial_\mu E_n(R,\mu)=E_n(R,\mu)\partial_R E_n(R,\mu)+\frac{1}{R}P_n(R)^2.
\end{align}
Now let us specify the undeformed QFT to be a conformal field theory, in which case we have $E_n\sim R^{-1}$ and $P_n\sim R^{-1}$. This allows us to solve the flow equation (\ref{eq:flowE}) which leads to
\begin{align}
	E_n(R,\mu)=\frac{R}{2\mu}\left(\sqrt{1+\frac{4\mu E_n}{R}+\frac{4\mu^2P_n^2}{R^2}}-1\right),
\end{align}
where $E_n$ and $P_n$ on the right hand side is the undeformed energy and momentum. It is convenient to introduce a dimensionless parameter $\lambda=-2\mu/(\pi R^2)$. Then the deformed energy can be written as
\begin{align}
	\label{eq:deformedE}
	\mathcal{E}(\lambda)=-\frac{1}{\pi \lambda R}	\left(  \sqrt{1-2\pi \lambda R E + \lambda^2 \pi^2 R^2 P^2} -1  \right)~,
\end{align}
where we have omitted the subscript `$n$' to simplify the notation.

As outlined in the introduction, we shall investigate the characters in the following double-scaling limit
\begin{align}\label{DS-limit}
	c\to \infty, \qquad
	\lambda \to 0, \qquad
 \tlambda\equiv 	\frac{\pi c\,\lambda}{6} =\text{fixed}~.
\end{align}
In addition, the analysis below will apply to characters of primaries in the conventional semi-classical regime of the CFT. That is, their conformal dimensions scale as
\begin{align}
	\Delta \to \infty, \qquad \frac{\Delta}{c} = \text{fixed}~.
\end{align}
\subsection{Method I: from energy spectrum}
To get things started, we shall consider deformed  characters\footnote{We adopt the terminology `deformed character' to refer to the $\ttb$ deformation of the product of left and right moving CFT characters of specific primary. At this point of time, the deformed symmetry algebra is unknown and it is unclear what the exact structure of an irrep in the deformed theory is. So we do not mean `character of an irrep' in a strict sense. Conservatively, we are evaluating the trace over a subset of the Hilbert space that smoothly flows to the CFT Verma module in the $\lambda\to 0$ limit.} of scalar primaries in a $c>1$ CFT with Virasoro as the chiral algebra. Before studying the deformed case, let us briefly recall the usual situation for CFT characters.
The left/right-moving character of given primary state is defined by the following trace over descendant states of the Verma module
\begin{align}
	\chi_h (\tau) =\tr_{\cal V}\! \left[e^{2\pi i \tau (L_0-c/24)}\right], \quad \chi_{\bar{h}} (\bar{\tau}) =\tr_{\overline{\cal V}}\! \left[e^{-2\pi i  \btau(\bL_0-c/24)}\right]~,
\end{align}
where, $\tau\equiv\tau_1+i\tau_2$ is the modular parameter of the torus.
The modular invariant partition function can then be written as a sum of characters
\begin{align}\label{char-sum}
	Z(\tau,\btau) = \sum_{h,\bar h} \chi_h (\tau) \chi_{\bar{h}} (\bar{\tau})~.
\end{align}

Let us parametrize the conformal dimension of the primary as follows
\begin{align}\label{delta}
	\Delta=h+\bh = \frac{c}{12}(1-\alpha^2)~.
\end{align}
The vacuum then corresponds to $\alpha=1$, while states above the BTZ threshold correspond to imaginary values of $\alpha$.
The descendants  of a scalar primary have the following energies and spins
\begin{align}\label{verma-energy}
	E R = - \frac{c\,\alpha^2}{12} +n_L + n_R~, \qquad P R = n_L-n_R~.
\end{align}
Here, $n_{L,R}$ are the left/right-moving descendant levels. For non-identity primaries, the degeneracy  at a given level $(n_L,n_R)$ is given by the product of integer partitions $p(n_L)p(n_R)$.
We can therefore write the product of left and right-moving characters as follows
\begin{align}\label{cft-char}
\Xi_\alpha(\tau,0)=	|\chi_\alpha(\tau)|^2 = e^{2\pi \tau_2 \frac{c\alpha^2}{12}} \sum_{n_L,n_R=0}^\infty p(n_L) p(n_R) e^{-2\pi \tau_2 (n_L+n_R)} e^{2\pi i\tau_1 (n_L-n_R)}~.
\end{align}
Here, we have defined the quantity $\Xi_\alpha(\tau,\tlambda)$ as the deformed product of left and right moving characters; the second argument $\tlambda$ denotes the $\ttb$ deformation parameter.
The summations over $n_L$ and $n_R$ can be performed separately and we can rewrite the above expression as
\begin{align}\label{cft-char-0}
\Xi_\alpha(\tau,0)= e^{2\pi \tau_2 \frac{c\alpha^2}{12}}
	\left|\prod_{n=1}^\infty {1\over 1-e^{2\pi i \tau n} } \right|^2, \qquad \tau=\tau_1 +i\tau_2~.
\end{align}
For the vacuum character, the above product starts out from $n=2$ accounting for the null states. Although all of this is known since antiquity, the details will turn out to be relevant for obtaining characters  of irrelevant deformations.

For $\ttb$ deformed CFTs, the energies are given \eqref{eq:deformedE}.
We focus on the double-scaling limit \eqref{DS-limit}. Additionally, we will work with $\lambda<0$ for now, so that we don't  run into complex energy issues.  In this regime of parameters, the deformed energies of the descendant states of a heavy primary state ($\Delta\sim O(c)$) are given by
\begin{align}\label{defE-1loop}
	\mathcal{E}
&= -\frac{c}{  6\tlambda }	\left(  \sqrt{1-12\tlambda\left(- \frac{\alpha^2 }{12} + \frac{n_L +n_R}{c}\right) + \frac{36 \tlambda^2}{c^2} (n_L-n_R)^2} -1  \right)\nn \\
&=	\frac{c \left(1-\sqrt{1+\alpha ^2 \tlambda }\right)}{6 \tlambda }+\frac{ {n_L+n_R}}{\sqrt{1+\alpha ^2 \tlambda }} +O(1/c)~.
\end{align}
In the second line we have expanded around $c\to\infty$ and kept terms till $O(c^0)$. Quite remarkably, up to this order the contribution from the descendant levels is linear just like the CFT spectrum \eqref{verma-energy}.
With this information in place, we can evaluate the following quantity
\begin{align}\label{ttb-char}
	\Xi_\alpha(\tau,\tlambda)~\approx~ e^{2\pi \tau_2 \frac{c \left(1-\sqrt{1+\alpha ^2 \tlambda }\right)}{6 \tlambda }} \sum_{n_L,n_R=0}^\infty p(n_L) p(n_R) e^{-2\pi \tau_2 \frac{(n_L+n_R)}{\sqrt{1+\alpha^2 \tlambda}}} e^{2\pi i\tau_1 (n_L-n_R)}~.
\end{align}
The `$\approx$' above (and below) indicates that $\log \Xi_\alpha$ has additional $O(1/c)$ corrections that we do not keep track of. The summations can be evaluated just like the CFT case and we obtain
\begin{align}\label{s-result}
	\Xi_\alpha(\tau,\tlambda)\approx e^{2\pi \tau_2 \frac{c \left(1-\sqrt{1+\alpha ^2 \tlambda }\right)}{6 \tlambda }}
	\left|\prod_{n=1}^\infty {1\over 1-e^{2\pi i \tau_1 n- 2\pi \frac{\tau_2}{\sqrt{1+\alpha^2 \tlambda}}n} } \right|^2.
\end{align}
This shows that, up to $O(1/c)$ the deformed character takes a similar form to that of a CFT.
We also observe that, at large $c$, the leading piece is simply the deformation of the conformal dimension/energy of the primary state -- the first line of \eqref{defE-1loop} with $n_{L,R}=0$. The next-to-leading-order   contribution takes an infinite-product form analogous to \eqref{cft-char}. Most importantly, the imaginary part of the modular parameter gets a rescaling that depends on the conformal dimension of the irrep. This is a dynamical change of coordinates. From the deformed energies of the descendants, we can deduce the commutation relations up to $O(1/c)$
\begin{align}
	[H,L_{-n}] \approx \frac{n}{\sqrt{1+\alpha^2\tlambda}}L_{-n}~, \qquad [H,\bar L_{-n}] \approx \frac{n}{\sqrt{1+\alpha^2\tlambda}}\bar L_{-n}~, \qquad \forall \, n \in \mathbb{Z}^+.
\end{align}
Note that the commutation relation depends on the conformal dimension of the primary state. This is in contrast to the CFT where the underlying symmetry algebra is itself responsible for classifying irreps.

The result \eqref{s-result} can also be derived from the deformed theory of reparametrization modes of AdS$_3$  \cite{Ouyang:2020rpq}.   Although we have considered deformation of Virasoro characters of non-vacuum scalar primaries  and $\lambda<0$ till this point, these features continue to hold more generally. We now turn to two other approaches that illustrate this point.

\subsection{Method II: integral kernel action}
The $\ttb$ deformation can be formulated as coupling the undeformed CFT to JT gravity. In this description, the deformed partition function on the torus can be obtained from the CFT one by the action of an integral kernel \cite{Dubovsky:2018bmo} (see also \cite{Hashimoto:2019wct})
\begin{align}
	Z_{\ttb}(\tau,\lambda) = - \frac{\tau_2}{\pi\lambda} \int_{\mathbb{H}^+} \frac{d^2 \zeta}{\zeta_2^2} e^{+\frac{1}{\lambda\zeta_2}|\zeta-\tau|^2} Z_{\rm cft}(\zeta)~.
\end{align}
The integral is over the upper-half  of the $\zeta$ plane.\footnote{For brevity, we mean the dependence on $(\tau,\btau,\lambda)$ when we write $(\tau,\lambda)$ and the dependence on $(\tau,\btau)$ when we write $(\tau)$.} If we write the CFT partition function as a sum over Boltzmann factors, then the kernel acts on each of the terms in the sum. Each integral then has the special feature of the semiclassical approximation (saddle point + quadratic fluctuations) being exact. This localization property is a consequence of the Duistermaat-Heckman theorem \cite{Dubovsky:2017cnj}.  We shall, however, express the partition function as a sum over characters \eqref{char-sum}. Each deformed character is then given by the same integral transformation
\begin{align}\label{int-kernel}
	\Xi_{\alpha}(\tau,\lambda) = - \frac{\tau_2}{\pi\lambda} \int_{\mathbb{H}^+} \frac{d^2 \zeta}{\zeta_2^2} e^{+\frac{1}{\lambda\zeta_2}|\zeta-\tau|^2} \Xi_\alpha(\zeta,0)~.
\end{align}
We now restrict attention to the double-scaling limit \eqref{DS-limit}. In order to proceed, let us separate the $O(c)$ and $O(c^0)$ contributions of the CFT character, cf.~\eqref{cft-char-0}
\begin{align}\label{cft-sep}
	\Xi_\alpha(\zeta,0) = e^{-cF_0(\zeta)}\,  \Xi^\ein_\alpha(\zeta,0)~,
\end{align}
where, $cF_0(\zeta)$ is the free energy at leading order.
Note that there are no approximations in the above equation and we are also allowing the possibility of arbitrary chiral algebras (super-Virasoro, $\cal W$ algebras, $U(1)^D$, etc.). However, there is one assumption: we are in an irrational CFT where the central charge is much larger than the number of conserved currents.   This is required for the $1/c$ expansion to make sense.

We now rescale $\lambda$ to  $\tlambda$ and substitute the factorized version of the character \eqref{cft-sep} in \eqref{int-kernel}\footnote{The integral \eqref{intt1} will converge only for $\tlambda<0$. We evaluate the integral in this safe domain and then analytically continue to $\tlambda>0$. The final result is seen to be consistent with the other approaches considered here. }
\begin{align}\label{intt1}
	\Xi_{\alpha}(\tau,\lambda) = - \frac{c\,\tau_2}{6\tlambda} \int_{\mathbb{H}^+} \frac{d^2 \zeta}{\zeta_2^2} e^{+\frac{\pi\, c }{6\tlambda\zeta_2}|\zeta-\tau|^2}e^{-cF_0(\zeta)}\,  \Xi^\ein_\alpha(\zeta,0)~.
\end{align}
In the large $c$ regime, we can evaluate the above integral by a saddle point approximation. The exponential part of the integrand on the right hand side of \eqref{intt1} can be written as $e^{c\,S_{\text{eff}}}$, where
\begin{align}
\label{eq:saddle-TTbar}
S_{\text{eff}}(\zeta_1,\zeta_2)=\frac{\pi}{6\tlambda\zeta_2}|\zeta-\tau|^2-F_0(\zeta)~.
\end{align}
The saddle for $(\zeta_1,\zeta_2)$ is at the solution of $\partial S_{\text{eff}}(\zeta_1,\zeta_2)/\partial\zeta_j=0$, $j=1,2$. Let us consider the vacuum module first, for which $F_0(\zeta) = -\frac{\pi \zeta_2}{6}$. Plugging into (\ref{eq:saddle-TTbar}), the saddle can be found readily
\begin{align}
	\zeta_1^* =\tau_1, \qquad \zeta_2^* = \frac{\tau_2}{\sqrt{1+\tlambda}}~.
\end{align}
The quadratic fluctuations about the saddle is given by the determinant of a $2\times 2$ Hessian matrix in $\zeta_{1,2}$ derivatives -- these cancel out exactly  with the prefactors in \eqref{intt1}. The final result is given by
\begin{align}\label{k-result-1}
	\Xi_1(\tau,\tlambda) \approx   e^{-\frac{\pi c\,\tau_2}{3\tlambda}(1-\sqrt{1+\tlambda})} \, \Xi^\ein_1\!\left(\tau_1,\frac{\tau_2}{\sqrt{1+\tlambda}}\right)~.
\end{align}
For non-identity heavy primaries, we have $F_0(\zeta) = -\frac{\pi \zeta_2}{6}\alpha^2$ (using the parametrization \eqref{delta}). The location of the saddle is then  at
\begin{align}
	\zeta_1^* =\tau_1, \qquad \zeta_2^* = \frac{\tau_2}{\sqrt{1+\alpha^2\tlambda}}~,
\end{align}
and the deformed characters read
\begin{align}\label{k-result-2}
	\Xi_\alpha(\tau,\tlambda) \approx   e^{-\frac{\pi c\,\tau_2}{3\tlambda}(1-\sqrt{1+\alpha^2\tlambda})} \, \Xi^\ein_\alpha\!\left(\tau_1,\frac{\tau_2}{\sqrt{1+\alpha^2\tlambda}}\right) ~.
\end{align}
This result is consistent with the deformation of Virasoro characters considered in the previous subsection, namely equation \eqref{s-result}. Once again, we see the state-dependent rescalings for the imaginary part of the modular parameter in the $O(c^0)$ piece. The analysis using the integral kernel can be generalized for characters of primaries with non-zero spin -- but we choose to tackle that using our next method.

\subsection{Method III: flow equation for the torus partition function}

Yet another means of examining the $\ttb$ deformed spectrum is via a diffusion-type equation for its partition functions. This PDE results from viewing the deformation as random fluctuations of the background metric \cite{Cardy:2018sdv} and also, independently, from modular bootstrap considerations \cite{Aharony:2018bad}. For the case of the torus, the PDE is
\begin{align}\label{pde0}
	-	\pd_\lambda Z =  \left[ \frac{\tau_2}{4}(\pd_{\tau_1}^2 +\pd_{\tau_2}^2 )+ \frac{\lambda}{2}\left(\pd_{\tau_2}-\frac{1}{\tau_2}\right)\pd_\lambda \right]Z~.
\end{align}
Each  Boltzmann factor, $e^{2\pi i[ \tau_1 P + i \tau_2 {\cal E}(\lambda) ]}$ with ${\cal E}(\lambda)$ given by \eqref{eq:deformedE}, is a solution to the above. As the PDE is linear, it implies that any linear combination of Boltzmann factors is also a solution. 
For our current purposes, we decompose the  partition function as a sum over characters and \eqref{pde0} is obeyed separately by each of the $\Xi_{\alpha,s}$'s -- the subscript $s$ indicates the spin of the primary. We rescale the coupling as before, $\tlambda=\pi c \lambda/6$, and have the following flow equation for the character
\begin{align}\label{flow-eq}
	-\frac{\pi c}{6}	\pd_{\tlambda}\, \Xi_{\alpha ,s}=  \left[ \frac{\tau_2}{4}(\pd_{\tau_1}^2 +\pd_{\tau_2}^2 )+ \frac{\tlambda}{2}\left(\pd_{\tau_2}-\frac{1}{\tau_2}\right)\pd_{\tlambda} \right]\Xi_{\alpha ,s} ~.
\end{align}
Note that in this format there is no restriction on the sign of $\lambda$, in contrast to the previous approaches. The initial condition at $\lambda=0$ is supplied by the CFT characters.

Our goal is to solve \eqref{flow-eq} in a $1/c$ expansion. To proceed, we consider the free energy, $F_{\alpha ,s}=-\log\Xi_{\alpha ,s}$, expanded around $c\to \infty$
\begin{align}\label{loop-exp-F}
	F_{\alpha ,s}(\tau,\tlambda)= \sum_{n=0}^\infty c^{1-n} F_{(\alpha,s),n} (\tau,\tlambda)~.
\end{align}
We shall keep the $\alpha$ and $s$ dependence implicit to lighten the notation.
The flow equation for the free energy is, from \eqref{flow-eq}
\begin{align}\label{flow-free}
	& 6\tau_2\tlambda\frac{\pd^2 F}{\pd \tau_2\pd\tlambda} +   3\tau_2^2 \left[    \frac{\pd^2F}{\pd \tau_1^2}+ \frac{\pd^2F}{\pd \tau_2^2}- c \left(\left(\frac{\pd F}{\pd \tau_1}\right)^2+ \left(\frac{\pd F}{\pd \tau_2}\right)^2\right) \right]\nn \\
	&+\frac{\pd F}{\pd \tlambda} \left(2\pi c \tau_2 - 6\tlambda - 6 c \tlambda \tau_2 \frac{\pd F}{\pd\tau_2}\right) =0 ~.
\end{align}
We now plug in the expansion \eqref{loop-exp-F} in the above and solve it order-by-order.

\subsubsection*{Scalar primary characters}
We consider the case with $s=0$ first.
At the zeroth order, $F_0$ only depends on $\tau_2$ and $\lambda$ and satisfies the following non-linear PDE
\begin{align}
	3 \tau_2 \left(\frac{\pd F_0}{\pd \tau_2}\right)^2 + 2 \frac{\pd F_0}{\pd \tlambda} \left( 3\tlambda \frac{\pd F_0}{\pd \tau_2}-\pi\right) =0 ~.
\end{align}
For scalar heavy primary states, the initial condition is $F_0(\lambda=0)= - \frac{\pi \tau_2}{6}\alpha^2$, using the parametrization \eqref{delta}. We have the solution
\begin{align}\label{F-tree}
	F_0 = -2\pi \tau_2 \left(\frac{\sqrt{1+\alpha^2\tlambda\,}-1}{6\tlambda}\right)~.
\end{align}
This is nothing other than the solution to the Burger's equation and is consistent with \eqref{s-result} and \eqref{k-result-2}. For the vacuum we can simply take $\alpha=1$ and we get the expected result, \ie the leading part of \eqref{k-result-1}.

At the next order, we have a differential equation that involves both $F_1$ and $F_0$. We can substitute the solution \eqref{F-tree} for $F_0$ and we obtain a \emph{linear} PDE for $F_1$
\begin{align}
	2(1+\alpha^2\tlambda) \frac{\pd F_1}{\pd\tlambda} + \alpha^2\tau_2 \frac{\pd F_1}{\pd \tau_2}  =0 ~,
\end{align}
with $\alpha=1$ for the vacuum module.
The $\tau_1$ dependence of $F_1$ remains unaffected. The most general solution to the above equation is any function of the form $f({\tau_2}/{\sqrt{1+\alpha^2 \tlambda}})$. Therefore, for the CFT initial conditions at $\tlambda=0$, we have the solution
\begin{align}\label{f1}
	F_1(\tau_1,\tau_2,\lambda) = F^{\rm (cft)}_1 \left(\tau_1,\frac {\tau_2}{\sqrt{1+\alpha^2 \tlambda}} \right)~.
\end{align}
Once again, this shows a state-dependent change of the modular parameter and agrees with the result \eqref{k-result-2} using the integral kernel.

\subsubsection*{Spinning primary characters}
We now consider deformation of characters of primaries with non-vanishing spin $s\neq 0$. In this case, we need to kept track of the $\tau_1$ dependence carefully. The zeroth order PDE reads
\begin{align}
	3 \tau_2 \left[\left(\frac{\pd F_0}{\pd \tau_1}\right)^2 +\left(\frac{\pd F_0}{\pd \tau_2}\right)^2 \right] + 2 \frac{\pd F_0}{\pd \tlambda} \left( 3\tlambda \frac{\pd F_0}{\pd \tau_2}-\pi\right) =0 ~.
\end{align}
The initial condition is given by $F_1(\tlambda=0)= 2\pi i \tau_1 S - \frac{\pi \tau_2 }{6}\alpha^2$ where $S\equiv s/c$. The solution is
\begin{align}\label{Fs-tree}
	F_0 = -2\pi i \tau_1 S-2\pi \tau_2 \left(\frac{\sqrt{1+\alpha^2\tlambda+ 36S^2 \tlambda^2\,}-1}{6\tlambda}\right)~.
\end{align}
At the next order, the PDE takes the form
\begin{align}\label{ord2}
	2G(\tlambda)^2 \frac{\pd F_1}{\pd\tlambda} + (\alpha^2+72 \tlambda S^2)\tau_2 \frac{\pd F_1}{\pd \tau_2} -12 i S\tau_2 G(\tlambda)\frac{\pd F_1}{\pd \tau_1}   =0 ~,
\end{align}
with $G(\tlambda) = \sqrt{1+ \alpha^2\tlambda+ 36 S^2\tlambda^2 }$. For the undeformed CFT, $F_1$ is a sum of holomorphic and anti-holomorphic parts
\begin{align}
	F_1(\tlambda=0) = \cF(\tau_1+i\tau_2)+ \bar\cF(\tau_1-i\tau_2)~.
\end{align}
The solution to \eqref{ord2} is then given by
\begin{align}
		F_1  = \cF\left(\tau_1+i\frac{\tau_2(1+6\tlambda S)}{\sqrt{1+ \alpha^2\tlambda+ 36 S^2\tlambda^2 }}\right)+  \bar\cF\left(\tau_1-i\frac{\tau_2(1-6\tlambda S)}{\sqrt{1+ \alpha^2\tlambda+ 36 S^2\tlambda^2 }}\right)~.
\end{align}
This reduces to the scalar primary case \eqref{f1} for $s=0$.
To illustrate this formulae, we write the deformed character of a spinning primary in a CFT with Virasoro symmetry
\begin{align}\label{flow-final}
	\Xi_{\alpha,s} (\tau,\tlambda)\approx\,& \exp\left[2\pi i \tau_1 s + \frac{\pi c\tau_2}{3\tlambda}\left(1-\sqrt{1+\alpha^2 \tlambda + 36 S^2 \tlambda^2 }\right)\right] \\
&\times  \left[\prod_{n=1}^{\infty}   \frac{1}{1-e^{2\pi i \tau_1 n - 2\pi  \frac{\tau_2(1+6S\tlambda)}{\sqrt{1+\tlambda(\alpha^2 + 36 \tlambda S^2)}} n  }}\right]\times  \left[\prod_{m=1}^{\infty}   \frac{1}{1-e^{-2\pi i \tau_1 m - 2\pi  \frac{\tau_2(1-6S\tlambda)}{\sqrt{1+\tlambda(\alpha^2 + 36 \tlambda S^2)}} m  }}  \right]  ~.\nn
\end{align}
The energy appearing in the leading exponential is simply the solution of the Burgers' equation for non-zero spin. The spin itself (which is the coefficient of $\tau_1$) remains unaffected owing to quantization on the spatial circle. At the next order, we observe that
the  $\tau_2$ in the left and right-moving sectors gets scaled differently.

\section[{Deformed characters of the $\ttb+\jtbar+T\bar{J}$ deformation}]{Deformed characters of the $\ttb+\jtbar+T\bar{J}$ \\ deformation}
\label{sec:joint}
The $T\bar{T}$ deformation has been generalized to more general solvable irrelevant deformations. We will derive the deformed character under these more general deformations in the large $c$ limit. To this end, it is most straightforward to apply the second method, namely the method of integration kernels.\par

For a theory with a holmorphic $U(1)$ current, one can define the $J\bar{T}$ deformation \cite{Guica:2017lia,Apolo:2018qpq,Chakraborty:2018vja,Anous:2019osb}. More generally, we can turn on several irrelevant deformations at the same time, and define the $\ttb+\jtbar+T\bar{J}$ deformation \cite{LeFloch:2019rut,Chakraborty:2019mdf}. The deformed spectrum and torus partition function have been studied in \cite{Aguilera-Damia:2019tpe,Chakraborty:2019mdf,Hashimoto:2019wct,Hashimoto:2019hqo,Chakraborty:2020xyz}. We denote the dimensionful deformation parameters by $\mu,\varepsilon_{\pm}$ such that the deformation is defined by
\begin{align}
\label{eq:genDef}
	\delta S=\int\left(\delta \mu\,T\bar{T}(x)+\delta\varepsilon_+\,J\bar{T}(x)+\delta\varepsilon_- T\bar{J}(x)\right)d^2x.
\end{align}
As before, putting the theory on a cylinder of radius $R$, we can introduce the dimensionless deformation parameters
\begin{align}
	\lambda=\frac{\mu}{R^2},\qquad \epsilon_+=\frac{\varepsilon_+}{R},\qquad \epsilon_-=\frac{\varepsilon_-}{R}.
\end{align}
For a state with momentum $P$, energy $E$ and the $U(1)$ charges $Q_L$ (holomorphic),$Q_R$ (anti-holomorphic), the deformed energy $\mathcal{E}$ is given by
\begin{align}\label{mixed-spec}
	R \mathcal{E} =  -
	\frac{1}{2\mathcal{A}}\left(\mathcal{B}-\sqrt{\mathcal{B}^2-4\mathcal{A}\mathcal{C}} \right)
\end{align}
where,
\begin{align}
\label{ak-def}
\mathcal{A}&= -1+h(\epsilon_+ -\epsilon_-)^2\kappa, \\\nonumber
\mathcal{B}&=2h s\kappa(\epsilon_-^2-\epsilon_+^2)-h(1-2Q_R\epsilon_-+2Q_L\epsilon_+),\\\nonumber
\mathcal{C}&=s^2\left(1+h\kappa(\epsilon_+ +\epsilon_-)^2\right)+2hs(Q_L\epsilon_+Q_R\epsilon_-)+h\Delta.
\end{align}
Here\footnote{Our normalization of $\lambda$ and $h$ differ slightly from the conventions of \cite{Hashimoto:2019wct}. The $h$ appearing in this section is not be confused with the holomorphic conformal dimension of the primary. }
\begin{align}
h = (-\tfrac{\pi}{2} \lambda+\tfrac{4}{\kappa}\ep_+\ep_-)^{-1}~,\qquad s=PR,\qquad \Delta=ER,
\end{align}
and $\kappa$ is level of the  $U(1)$ current algebra.

Analogous to the $\ttb$ case, the torus partition function under the deformation (\ref{eq:genDef}) can be obtained by an integral transformation. The integration kernel can be derived by realizing the deformation as coupling the seed CFT to a 2d gauge theory and a JT-like topological gravity \cite{Aguilera-Damia:2019tpe}, and then performing the path integral on the torus.
Yet another method to arrive at the integral kernel is  by utilizing the string theory dual of the deformation with general couplings \cite{Chakraborty:2019mdf,Hashimoto:2019wct}. The deformed partition function is given by\footnote{Analogous to the previous section, we write the dependence on $(\tau,\btau,\lambda,\ep_\pm)$ as $(\tau,\lambda,\ep_\pm)$ and the dependence on $(\tau,\btau,\chi,\bar\chi)$ as $(\tau,\chi)$.}
\begin{align}
	Z(\tau,\lambda,\ep_\pm) &= \int_{\mathbb{H}^+} d^2 \zeta \int_{\mathbb{C}} d^2 \chi ~{\cal K} (\tau,\zeta,\chi) \, Z_{\rm inv}(\zeta, \chi)~ , \\
	{\cal K} (\tau,\zeta,\chi) &= \frac{\tau_2}{4\ep_+\ep_-\zeta_2^3} \exp \left[-\frac{\pi \chi\bar\chi}{2h \ep_+\ep_-\zeta_2} - \frac{\pi\chi}{2\ep_+\zeta_2} (\bar\zeta-\btau)  + \frac{\pi\bar\chi}{2\ep_-\zeta_2} (\zeta-\tau)
	\right]~.\nn
\end{align}
The kernel acts on the following modular invariant combination of the flavored partition function of the CFT and an exponential factor involving the $U(1)$ chemical potentials
\begin{align}\label{gen-kernel}
	Z_{\rm inv}(\tau, \chi)~=~e^{\pi\kappa \frac{ (\chi-\bar\chi)^2}{2\tau_2}}Z_{\rm cft}(\tau, \chi)~=~e^{\pi\kappa \frac{ (\chi-\bar\chi)^2}{2\tau_2}}
	\tr\!\left[q^{L_0-c/24}\bq^{\bL_0-c/24} e^{2\pi i \chi J_0}e^{-2\pi i \bar\chi \bar J_0} \right]~,
\end{align}
where, $q=e^{2\pi i \tau}$.

We now consider the double-scaling limit of the deformed partition function. More precisely, we take the limit
\begin{align}
	c\to \infty, \qquad \lambda\to 0, \qquad \ep_\pm\to 0~.
\end{align}
while keeping the following rescaled couplings fixed
\begin{align}
	\tlambda = \frac{\pi c\,\lambda}{6}, \qquad \tilde{\epsilon}_\pm = c\,\ep_\pm~.
\end{align}
 Note that the level of the current algebra  scales with the central charge\footnote{For $N$ complex free fermions $\kappa=N=c$, for $N$ free bosons $\kappa=N=c$ and for $\cN=2$ SCFTs $\kappa=c/3$.} and this motivates defining the quantities $k=\kappa/c$ and $\tilh = (3\tlambda+  k \tep_+ \tep_-)^{-1}=h/c $. Using these definitions we can write the deformed characters as
\begin{align}
\label{gen-kernel-2}
	\tilde\Xi_{\alpha,q_{L,R}}(\tau,\lambda,\ep_\pm) &= \frac{\tau_2}{4\ep_+\ep_-} \int_{\mathbb{H}^+}\! \frac{d^2 \zeta}{\zeta_2^3} \int_{\mathbb{C}} d^2 \chi ~ e^ {c\left[-\frac{\pi \chi\bar\chi}{2\tilh \tilde{\epsilon}_+\tilde{\epsilon}_-\zeta_2} - \frac{\pi\chi}{2\tilde{\epsilon}_+\zeta_2} (\bar\zeta-\btau)  + \frac{\pi\bar\chi}{2\tilde{\epsilon}_-\zeta_2} (\zeta-\tau)
		\right]} \, e^{\pi k c \frac{ (\chi-\bar\chi)^2}{2\zeta_2}}\Xi^{\rm cft}_{\alpha,q_{L,R}}(\zeta, \chi)~ .
\end{align}
The symbols appearing above deserve an explanation.
The CFT character within the integral is in the grand canonical ensemble, with non-zero chemical potentials corresponding to the $U(1)$ charges,
\begin{align}
\Xi^{\rm cft}_{\alpha,q_{L,R}}(\zeta,\chi) = 	\tr_{\cal V}\!\left[q^{L_0-c/24}e^{2\pi i\chi J_0}\right] \tr_{\overline{\cal V}}\!\left[\bq^{\bL_0-c/24}e^{-2\pi i\bar\chi\bar J_0}\right]~.
\end{align}
On the other hand, the deformed character on the LHS have zero chemical potentials but non-vanishing irrelevant couplings.
We shall consider the case of charged  primaries, with non-zero spin, for which the product of left and right moving CFT characters is 
\begin{align}\label{f-cft-sep}
	\Xi^{\rm cft}_{\alpha,q_{L,R}}(\zeta,\chi) = e^{2c\pi i\zeta_1\tilde{p}+c\zeta_2\frac{\pi\alpha^2}{6}}\, e^{2\pi i  \chi c q_L}e^{-2\pi i\bar\chi c q_R}\,  \Xi^\ein_{\alpha,q_{L,R}}(\zeta,\chi)~.
\end{align}
The conformal dimension is parametrized as before in \eqref{delta} and we have rescaled the spin of the primary state as $PR=c\,\tilde{p}$, as well as the charges of the primary as $Q_{L,R}=c\, q_{L,R}$ --- that is, we are in the sector of large spin and charges. This is the most general case in the large $c$ limit. 
 $\Xi^\ein_{\alpha,q_{L,R}}(\zeta,\chi)$ is the $O(c^0)$ part of the free energy and it contains the contribution from the descendants. The functional dependence on $\tau$ and $\chi$ depends on the chiral algebra of the CFT. The only restriction we impose here is that both the left and right chiral algebra should contain an $U(1)$ current.

We can substitute \eqref{f-cft-sep} in \eqref{gen-kernel-2} to obtain the deformed character.  As we are interested in the large $c$ regime, the integral can be evaluated in the saddle-point approximation.\footnote{It is profitable to work with real integration variables -- we perform the change, $\chi=\chi_1+i{\chi_2}$ and $\bar\chi=\chi_1-i {\chi_2}$, and then locate the saddle.} To this end, it is convenient to rewrite (\ref{gen-kernel-2}) as
\begin{align}
\tilde\Xi_{\alpha,q_{L,R}}(\tau,\lambda,\ep_\pm) &= \frac{\tau_2}{4\ep_+\ep_-} \int_{\mathbb{H}^+}\! \frac{d^2 \zeta}{\zeta_2^3} \int_{\mathbb{C}} d^2 \chi ~ e^ {c\,S_{\text{eff}}}\,\Xi^{(1)}_{\alpha,q_{L,R}}(\zeta, \chi)~,
\end{align}
where $S_{\text{eff}}$ is given by
\begin{align}
\label{eq:Seffdef}
S_{\text{eff}}=&\,\frac{k\pi(\chi-\bar{\chi})^2}{2\zeta_2}-\frac{\pi\chi\bar{\chi}}{2\tilde{h}\tilde{\epsilon}_+\tilde{\epsilon}_-\zeta_2}- \frac{\pi\chi}{2\tilde{\epsilon}_+\zeta_2} (\bar\zeta-\btau)  + \frac{\pi\bar\chi}{2\tilde{\epsilon}_-\zeta_2} (\zeta-\tau)\\\nonumber
&\,+2\pi i\zeta_1\tilde{p}+\zeta_2\frac{\pi\alpha^2}{6}+2\pi i  \chi q_L-2\pi i\bar\chi q_R~.
\end{align}
In the large $c$ limit, the contribution is dominated by the saddle-point
\begin{align}
\frac{\partial S_{\text{eff}}}{\partial\chi}=\frac{\partial S_{\text{eff}}}{\partial\bar{\chi}}=\frac{\partial S_{\text{eff}}}{\partial\zeta_1}=\frac{\partial S_{\text{eff}}}{\partial\zeta_2}=0~.
\end{align}
The solution can be found straightforwardly. After some manipulation, it reads
\begin{align}
\label{saddle-val}
&\chi^*=\frac{i\tilde{h}\tilde{\epsilon}_+\tau_2}{\tilde{h}\tilde{\kappa}(\tilde{\epsilon}_- -\tilde{\epsilon}_+)^2-1 }\left(1+\frac{\sqrt{3}(\mathbb{C}_1+2\tilde{p})}{\sqrt{\mathbb{A}}}\right),\\\nonumber
&\bar{\chi}^*=\frac{i\tilde{h}\tilde{\epsilon}_-\tau_2}{\tilde{h}\tilde{\kappa}(\tilde{\epsilon}_- -\tilde{\epsilon}_+)^2-1 }\left(1+\frac{\sqrt{3}(\mathbb{C}_2-2\tilde{p})}{\sqrt{\mathbb{A}}}\right),\\\nonumber
&\zeta_1^*=\tau_1-\frac{i\tau_2}{\tilde{h}\tilde{\kappa}(\tilde{\epsilon}_- -\tilde{\epsilon}_+)^2-1}\left(\tilde{h}\tilde{\kappa}(\tilde{\epsilon}_-^2-\tilde{\epsilon}_+^2)-\frac{\sqrt{3}\mathbb{C}_3}{\sqrt{\mathbb{A}}} \right),\\\nonumber
&\zeta_2^*=-\frac{\sqrt{3}\tilde{h}}{\sqrt{\mathbb{A}}}\tau_2~,
\end{align}
where,  the following objects are defined to shorten the expressions
\begin{align}
\mathbb{A}=&\,12\tilde{p}^2(1+4\tilde{h}\tilde{\kappa}\tilde{\epsilon}_+\tilde{\epsilon}_-)\\\nonumber
&\,+12\tilde{p}\tilde{h}\left[2q_R\tilde{\epsilon}_-+2q_L\tilde{\epsilon}_+ - \tilde{h}\tilde{\kappa}(\tilde{\epsilon}_- -\tilde{\epsilon}_+)
(\tilde{\epsilon}_-+\tilde{\epsilon}_++4(q_L-q_R)\tilde{\epsilon}_+\tilde{\epsilon}_-)\right]\\\nonumber
&\,+\tilde{h}\left[3\tilde{h}(1-2q_R\tilde{\epsilon}_-+2q_L\tilde{\epsilon}_+)^2
+\alpha^2(\tilde{h}\tilde{\kappa}(\tilde{\epsilon}_--\tilde{\epsilon}_+)^2-1) \right]~,
\end{align}
and
\begin{align}
\mathbb{C}_1=&\,4\tilde{h}\tilde{p}(\tilde{\epsilon}_+ -\tilde{\epsilon}_-)\tilde{\epsilon}_-\tilde{\kappa}+\tilde{h}(1-2q_R\tilde{\epsilon}_-+2q_L\tilde{\epsilon}_+),\\\nonumber
\mathbb{C}_2=&\,4\tilde{h}\tilde{p}(\tilde{\epsilon}_+ -\tilde{\epsilon}_-)\tilde{\epsilon}_+\tilde{\kappa}+\tilde{h}(1-2q_R\tilde{\epsilon}_-+2q_L\tilde{\epsilon}_+),\\\nonumber
\mathbb{C}_3=&\,2\tilde{p}(1+4\tilde{h}\tilde{\epsilon}_-\tilde{\epsilon}_+\tilde{\kappa})
+\tilde{h}\left[2q_R\tilde{\epsilon}_-+2q_L\tilde{\epsilon}_+ - \tilde{h}\tilde{\kappa}(\tilde{\epsilon}_- -\tilde{\epsilon}_+)
(\tilde{\epsilon}_-+\tilde{\epsilon}_++4(q_L-q_R)\tilde{\epsilon}_+\tilde{\epsilon}_-)\right].
\end{align}
Plugging the saddle-point solution \eqref{saddle-val} into (\ref{eq:Seffdef}), we obtain the `on-shell' effective action
\begin{align}
S_{\text{eff}}^*=2\pi i\tau_1\,\tilde{p}-
	\frac{\pi \tau_2\left(\sqrt{\mathbb{A}/3}+\tilde{h}(1-2q_R\tilde{\epsilon}_-
+2q_L\tilde{\epsilon}_++2\tilde{p}(\tilde{\epsilon}_+^2-\tilde{\epsilon}_1^2)\tilde{\kappa})\right)}
{\tilde{h}(\tilde{\epsilon}_--\tilde{\epsilon}_+)^2\tilde{\kappa}-1}~.
\end{align}
The quadratic fluctuations about the saddle point can also be evaluated; this is the determinant of the 4-dimensional Hessian matrix that contains quadratic derivatives in $(\tau_1, \tau_2, \chi_1, \chi_2)$. The final result for the deformed character is
\begin{align}\label{mix}
	\tilde\Xi_{\alpha,q_{L,R}}(\tau,\lambda,\ep_\pm) \approx  e^{c\,S_{\text{eff}}^*}\,\Xi^\ein _{\alpha,q_{L,R}}(\zeta^*,\chi^*)~.
\end{align}
with $\zeta^*,\chi^*$ being the saddle point values \eqref{saddle-val}.
For the case of $\tilde{\ep}_\pm=0,\tilde{p}=0$, we have $\tilh=-(3\tlambda)^{-1}$ and the above expression agrees with the $\ttb$ deformed result \eqref{k-result-2}.

 Analogous to the $\ttb$ case we observe that the leading $O(c)$ part of \eqref{mix} is simply given by the deformed energy \eqref{mixed-spec} of the charged primary state, with unchanged spin. The $O(c^0)$ contribution is the same as that of the flavored characters, but with  modular parameters and $U(1)$ chemical potentials given by the saddle point values. Once again, we get primary-state-dependent rescalings for the modular parameter. The $U(1)$ chemical potentials become functions of the modular parameter, the irrelevant couplings as well as the conformal dimension and the spin of the primary.

\section{Conclusions and discussions on holography}
\label{sec:conclusions}

In this work, we studied a double-scaling limit of $\ttb$ deformed  characters in the $1/c$ expansion. Our main finding is that the  $O(c^0)$ contribution is deformed in a surprisingly simple way --- by a state-dependent rescaling  of the modular parameter. We arrived at this result by three approaches, namely, directly using the deformed spectrum, the integral kernel for deformed partition functions and the flow equation. These methods are equivalent and they give consistent results.  We also studied the more general $\ttb+\jtbar+T\bar J$ deformation and found a similar behavior for the characters.

Unlike the case of irrational CFTs, the $1/c$ expansion of the logarithm of deformed characters does not truncate. It would be interesting to explore the structure of deformed characters at higher orders in this expansion. A systematic way to proceed is to use the flow equation for the free energy, equation \eqref{flow-free}.
Furthermore, the $\ttb$ deformed partition function has well-defined modular properties \cite{Datta:2018thy,Aharony:2018bad}. It would be of value to use the results for the characters to find the fusion kernel for the S-modular transformation. Note that the dependence on the conformal dimension of the primary operator enters in a non-trivial manner in the deformed characters, which might lead to some intricacies.

It is also worthwhile to explore  partition functions of $\ttb$ deformed CFTs with boundaries using techniques similar to this work. In particular, the boundary states of Liouville theory -- FZZT branes -- can be analyzed. For a given boundary state, the undeformed partition function is simply given the left-moving character. The manner in which this quantity gets deformed can be determined by the flow equation for the finite cylinder geometry \cite{Cardy:2018sdv}. More fundamentally, it would be crucial to understand how the boundary states themselves get deformed. Since the $\ttb$ operator manifestly couples the left and right-moving sectors, the structure of a boundary state is expected to be quite different from the CFT case.

Much of the present work has been motivated by the holographic proposals of the $\ttb$ deformation \cite{McGough:2016lol,Kraus:2018xrn,Guica:2019nzm,Hirano:2020nwq}. So far, most of the checks for holography have been done in the classical limit, namely at the leading order in $1/c$ expansion (see the very recent work \cite{Kraus:2021cwf} for an exception). In order to extend the $T\bar{T}$ holographic dictionary to the quantum regime, it is important to test the $1/c$ corrections on both sides. The deformed characters we obtained in this paper offer concrete and simple predictions for the bulk, which can be tested in  future. Let us spell out these predictions in more detail.

We consider the deformed Virasoro characters which was derived in Section \ref{sec:ttb}. For a scalar primary, we have
\begin{align}\label{s-result-2}
	\Xi_\alpha(\tau,\tlambda)\approx e^{2\pi \tau_2 \frac{c \left(1-\sqrt{1+\alpha ^2 \tlambda }\right)}{6 \tlambda }}
	\left|\prod_{n=1}^\infty {1\over 1-e^{2\pi i \tau_1 n- 2\pi  \frac{\tau_2}{\sqrt{1+\alpha^2 \tlambda}}n} } \right|^2.
\end{align}
For the case of the vacuum module, we need to set $\alpha=1$ and the infinite product starts out with $n=2$. The large central charge regime of 2d CFTs corresponds to the semiclassical regime of gravity in AdS$_3$; this is implied by the Brown-Henneaux relation $c=\frac{3\ell_{\rm AdS}}{2G_N}$. The leading-order/tree-level part can be reproduced by the holographic proposals for $\ttb$ -- these arise from the on-shell gravitational action of a given background geometry, supplemented with appropriate boundary conditions to take the deformation into account.
The more interesting piece is the $1/c$ contribution of these characters. These corresponds to one-loop determinants on the gravity side. To be specific, we have the following expectations for one-loop determinants of the graviton in thermal AdS$_3$, BTZ black hole and conical defect (with a deficit angle $2\pi (1-\alpha)$) geometries:
\begin{align}\label{1-loop-expectations}
	\Xi^\ein_{\rm AdS_3} = \left|\prod_{n=2}^\infty {1\over 1-e^{-   \frac{2\pi\beta/R}{\sqrt{1+ \tlambda}}n} } \right|^2\!,~  \Xi^\ein_{\rm BTZ} = \left|\prod_{n=2}^\infty \frac{1}{1-e^{ -   \frac{2\pi R}{\sqrt{\beta^2+\tlambda R^2}} n  }}\right|^2\!, ~
	\Xi^\ein_{\alpha} = \left|\prod_{n=1}^\infty {1\over 1-e^{-   \frac{2\pi\beta/R}{\sqrt{1+ \alpha^2\tlambda}}n} } \right|^2\!.
\end{align}
For the BTZ case, we used the S-modular transformation of the thermal AdS$_3$ result; the deformation parameter transforms as $\tlambda\mapsto \tlambda/|\tau|^{-2}$ \cite{Datta:2018thy}.    In the above relations,  we have taken the boundary torus to be rectangular with the modular parameter $\tau=i\beta/R$. Note that unlike the case of pure gravity, the deformed characters are not one-loop exact, \ie the $1/c$ expansion of $\log \Xi_\alpha$ does not truncate.

There are various approaches for evaluating one-loop determinants in AdS -- these include the Selberg trace formula \cite{Bytsenko:1997sr,Bytsenko:1997ru}, heat-kernel methods \cite{Mann:1996ze,Giombi:2008vd,David:2009xg}, covariant phase space methods \cite{Cotler:2018zff}  and the method of (quasi)normal modes \cite{Denef:2009kn}. The recent paper \cite{Kraus:2021cwf} achieves the objective of obtaining the one-loop contribution using  covariant phase space methods adapted to cutoff AdS$_3$; this reproduces the expectation for thermal AdS$_3$, $\Xi^\ein_{{\rm AdS}_3}$ in \eqref{1-loop-expectations}. The (quasi)normal mode method also seems to be well suited for obtaining the one-loop determinant in the present context. In this approach, the determinant of the Laplacian is expressed as a product over normal modes for geometries without horizons, such as thermal AdS$_3$, and quasinormal modes (QNMs) for black holes.   For the BTZ black hole, we observe from \eqref{1-loop-expectations} that the  Hawking temperature is modified as follows
\begin{align}\label{temp-shift}
	\beta \mapsto \hat\beta\equiv\sqrt{\beta^2 + \tlambda R^2}~.
\end{align}
In the cutoff AdS proposal, the combination $\tlambda R^2$ appearing in \eqref{temp-shift} is related to the cutoff radius as \cite[eq.\,(2.17) and (3.7)]{McGough:2016lol}
\begin{align}
	\tlambda R^2 = \frac{\pi c\, \mu}{6} = \frac{1}{r_c^2}~.
\end{align}
The QNMs of the graviton in BTZ were evaluated in \cite{Datta:2011za} and were used to construct the one-loop determinant (see also \cite{Castro:2017mfj}).
As the form of the determinant remains the same, it implies that the QNMs also retain their original form, albeit the temperature shift \eqref{temp-shift}. The modes are expected to be the following
\begin{align}\label{qnms}
	\omega_L = k - 2\pi i  \hat\beta^{-1} (2n+\hat\Delta-1)~, \quad \omega_R = -k - 2\pi i  \hat\beta^{-1} (2n+\hat\Delta+1)~,
\end{align}
with $\hat\Delta=1+m$ and $n=0,1,2,\cdots$.\footnote{In \eqref{qnms} the graviton is massive. To evaluate the 1-loop determinant for the massless graviton, one has to fix gauge redundancies which is tantamount to including a ghost determinant and choosing the masses: $m^2=1$ for the graviton and $m^2=4$ for the ghost.} This prediction for the QNMs should be reproduced by the holographic constructions for the $\ttb$ deformation, otherwise appropriate modifications are necessary.

\section*{Acknowledgements}
We thank Per Kraus for discussions and helpful comments on the draft. 
We also thank Pawel Caputa, Soumangsu Chakraborty and Arunabha Saha for fruitful discussions. 
\begin{small}

\providecommand{\href}[2]{#2}

\end{small}

\end{document}